# Creep Behaviour and Tensile Response of Adhesively Bonded Polyethylene Joints: Single-Lap and Double-Strap

M.A. Saeimi Sadigh [a], B. Paygozar [b,*], L.F.M. da Silva [c], E. Martínez-Pañeda [d]

[a] Department of Mechanical Engineering, Azarbaijan Shahid Madani University, Tabriz, Iran

[b] Department of Mechanical Engineering, TED University, Ankara, Turkey

[c] Department of Mechanical Engineering, Faculty of Engineering, University of Porto, Porto, Portugal

[d] Department of Civil and Environmental Engineering, Imperial College London, London SW7 2AZ, UK

* Corresponding: Saeimi.sadigh@azaruniv.ac.ir

**Abstract:** The static and time-dependent behaviours of adhesively bonded polyethylene Double-Strap (DS) joints were investigated to assess the viability of this joint configuration relative to the Single-Lap (SL) joints. Both experiments and finite element simulations are conducted. First, we individually characterise the tensile and creep behaviour of the adhesive and adherent materials; an epoxy-based adhesive and polyethylene, respectively. This information is used to develop suitable constitutive models that are then implemented in the commercial finite element package ABAQUS by means of user material subroutines, UMATs. The numerical models are used to design the creep tests on the adhesive joints. Afterwards, an extensive experimental campaign is conducted where we characterise the static and creep behaviour of two joint configurations, SL and DS joints, and three selected values of the overlap length. In regard to the static case, results reveal an increase in the failure load with increasing overlap length, of up to 10% for an overlap length of 39 mm. Also, slightly better performance is observed for the SL joint configuration. For the creep experiments, we show that the DS adhesive joint configuration leads to much shorter elongations, relative to the SL joints. These differences diminish with increasing overlap length but remain substantial in all cases. In both joint configurations, the elongation increases with decreasing overlap length. For instance, increasing the overlap length to 39 mm led to a 50% and a 30% reduction in elongation for SL and DS joints, respectively. Moreover, the numerical predictions show a good agreement with the experiments. The stress redistribution is investigated and it is found that the shear stress is highly sensitive to the testing time, with differences being more noticeable for the DS joint system. The findings bring insight into the creep behaviour of polyethylene-based adhesive joints, a configuration of notable industrial interest.

**Keywords:** Creep behaviour; double-strap joint; single-lap joints; polyethylene.

# 1. Introduction

Adhesive joints are widely used across many sectors due to their advantages relative to other competing joining technologies. Well-known advantages include weight reduction, fewer sources of stress concentration and reduced through-life maintenance [1]. Adhesive joints can be classified into multiple groups according to their configuration; these include Single-Lap (SL), double-lap, and scarf joints. In recent years, there has been an increasing interest in the repair of adhesive joints, often using straps. This is motivated by applications exposed to potential sources of damage, especially in aeronautics, where other joining techniques such as riveting or bolting are not an option. As a consequence, a burgeoning literature has emerged with the aim of mitigating subsequent damage due to cracking (see, e.g. [2] and references therein). This is often achieved by optimising the stress distribution. For example, by reducing the stiffness at the ends of the overlap, tapering the surface of the patches or using fillets filled with adhesives [3]. Temiz modified the stress distribution in the adhesive layers by means of the mixed modulus joint concept [4]. Also, embedded patches can be used to alter the stress distribution and augment the load transfer capacity of the joint. Campilho *et al.* [5] investigated the tensile behaviour of adhesive single and Double-Strap (DS) joints with carbon-epoxy substrates. They found the optimal overlap length to be of 15 mm and the repair strength to be insensitive to changes in patch thickness. However, when buckling load is considered the optimal overlap length changes and a sensitivity of repair strength to patch thickness is revealed by Campilho et. al [6].

A promising material for joints adherents is polyethylene (PE), widely used in a variety of sectors. Polyethylene can favourably compete with metals due to its higher strength-to-weight ratio, bonding performance and resistance against corrosion [7]. Studies have been conducted regarding the use of polyethylene adherents in lap-shear joints. Pinto *et al.* [8] measured the bonding strength of SL joints of adherents made of polyethylene, composite and aluminium. They found a high sensitivity to the surface preparation technique. The influence of surface preparation was also assessed by Barton and Birkett [9], who investigated the tensile strength and impact behaviour of SL joints with PE adherents. Also in the context of SL joints, LeBono *et al.* [10] investigated the lap-shear strength performance of polyethylene pipeline bonded with an acrylic adhesive in the temperature range -10 to +20 °C. They found that a decrease in curing/testing temperature to zero degrees resulted in a steady reduction in the lap-shear strength performance of the bonded joints. Recently, Dehaghani *et al.* [11] assessed the influence of acid etching duration on the adhesive bonding strength of polyethylene on E-glass/epoxy composites. They found that both joint strength and fatigue resistance improved with increasing acid etching exposure time.

The response of adhesive joints under creep is of interest to many applications across the aerospace, transport, energy, and marine sectors [12]. Accordingly, a number of experimental and numerical studies have been devoted to characterising the creep behaviour of adhesive joints. For instance, Dean [13] developed a model for non-linear creep in an epoxy adhesive under both dry and humid conditions. Yu *et al.* [14] investigated the rate-dependent behaviour of epoxy-based adhesives using both power-law creep models and so-called unified theory models. Saeimi Sadigh *et al.* [15] combined experiments and modelling to characterise the creep behaviour of epoxy-based adhesive joints at different temperature levels. However, none of these studies deals with adhesively bonded polyethylene joints, motivating the present study. The creep behaviour of polyethylene has been a subject of interest outside of the adhesive joint community [16]. We build on this knowledge to characterise the behaviour of polyethylene-based adhesive joints.

In this work, we investigate for the first time the creep behaviour of adhesively bonded polyethylene strap joints. Both experiments and finite element simulations are conducted. First, the tensile and creep behaviours of the adherent and adhesive materials are characterised, experimentally and numerically. This information is then used to assess the performance of SL and DS joint configurations. Comparisons are drawn between the behaviour exhibited by SL and DS joints. However, the aim is not to compare performances but to characterise, numerically and experimentally, the behaviour of various classes of adhesive joints based on polyethylene. The different conditions present in SL and DS joint configurations enable assessing the generality of the constitutive models developed. Tensile tests are first conducted to characterise the mechanical response of the adhesive joints under monotonic loading. Creep tests are then performed to gain insight into the time-dependent response. In all cases, the role of the adherent length is explored by testing three different cases per adhesive joint configuration. In addition, finite element analysis is also conducted to gain further insight and assist in the interpretation of the results. The validation of the numerical model gives confidence in the use of parametric finite element analysis studies for joint design across many applications.

The paper is organised as follows. Section 2 presents the details of the experimental campaign. In Section 3, we describe the constitutive material models employed and the finite element framework developed. The numerical and experimental results are presented and discussed in Section 4. Finally, the manuscript conclusions are presented in Section 5.

## 2. Experimental study

We proceed to describe the testing procedures under tensile and uniaxial creep loading conditions. Tests were carried out on the adherent and adhesive materials, individually, and on both Single-Lap (SL) and Double-Strap (DS) joints, see Fig. 1. The outcome of the experimental campaign was used to validate the finite element model. The adherents of both the SL and DL joints were made of 5 mm thick high-density polyethylene (HDPE) sheets with dimensions 25x120 mm. These are bonded through a thin layer (0.2 mm) of epoxy-based structural adhesive, Araldite 2011 (Huntsman Advanced Materials, India). The mechanical properties of HDPE and the adhesive, as well as their time-dependent behaviour, are characterised by testing dog-bone samples manufactured following ASTM D638 standard, as shown in Fig. 1b.

Two batches of SL and DS joints are manufactured with geometries and dimensions as given in Fig. 1c and 1d. Three different configurations were considered in each group, with characteristic overlap lengths of $l_a$=19 mm, 29 mm and 39 mm. Surface preparation of the adherents is carried out prior to the bonding process. Following ASTM D 1780 standard recommendations, abrasive sandpaper is used to ensure a rough surface in the bonding domain of the joints. Subsequently, the adherents' surfaces were cleaned with acetone to remove contamination, followed by an etching process through $H_2SO_4$ solution. During this process, the solution removes the remaining organic matters from the substrates and hydroxylates them. Consequently, the surfaces will be more hydrophilic. A special fixture was used to ensure alignment of the adherents during the production process.

We examined first the tensile and creep behaviours of the adherent and adhesive materials. The uniaxial tension tests were conducted with a crosshead speed of 1.3 mm/min. While for the creep testing, we used constant loads at 65%, 75% and 85% of the joint strength (at room temperature). For this purpose, an automatic creep testing machine was employed to record the time-dependent displacement of the samples by means of an extensometer with 0.02 mm accuracy. The loading arm of the machine adjusts the horizontal position automatically to ensure the application of a constant load during the test. The outcome of the tests was then used to inform the modelling of static and creep behaviours by means of suitable constitutive models.

Regarding the mechanical behaviour of the adhesive joints, both uniaxial tension tests and creep experiments were conducted. Tension tests are used to measure the load-displacement response and their maximum strength (Fig. 1a), both relevant parameters in adhesive joint design. The loading rate corresponds to that used for the uniaxial tests on the adherent and the adhesive materials. Afterwards, creep tests were conducted to characterise the time-dependent behaviour of the joints. Creep tests of SL and DS joints were carried out at constant loads of 1.6 and 2 kN, and at a constant temperature of 25°C.

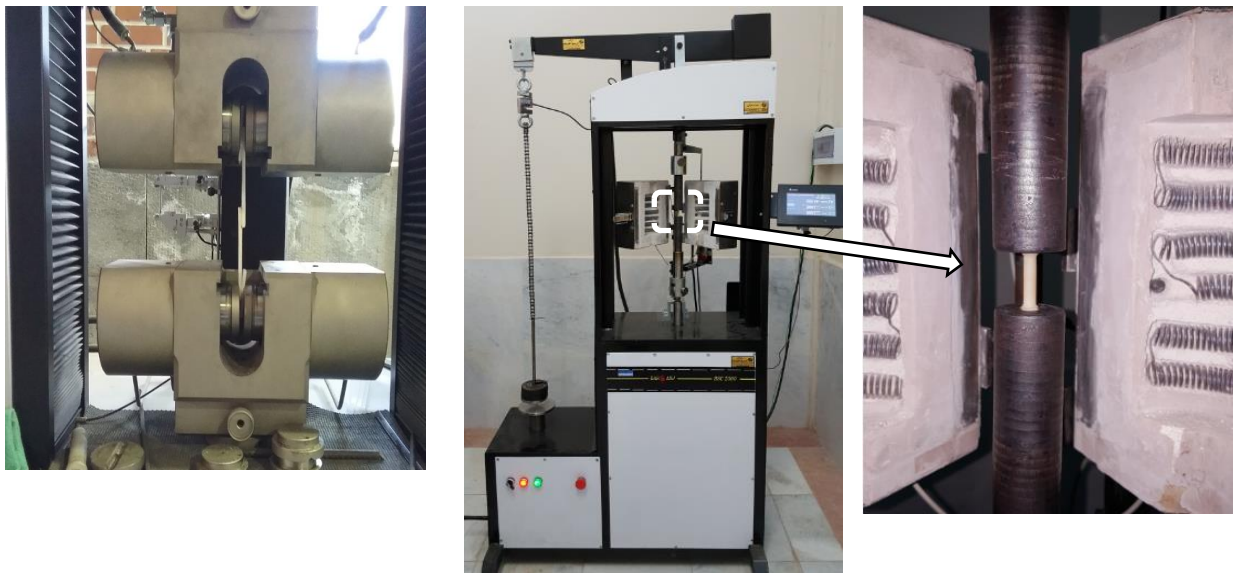

(a) (b)

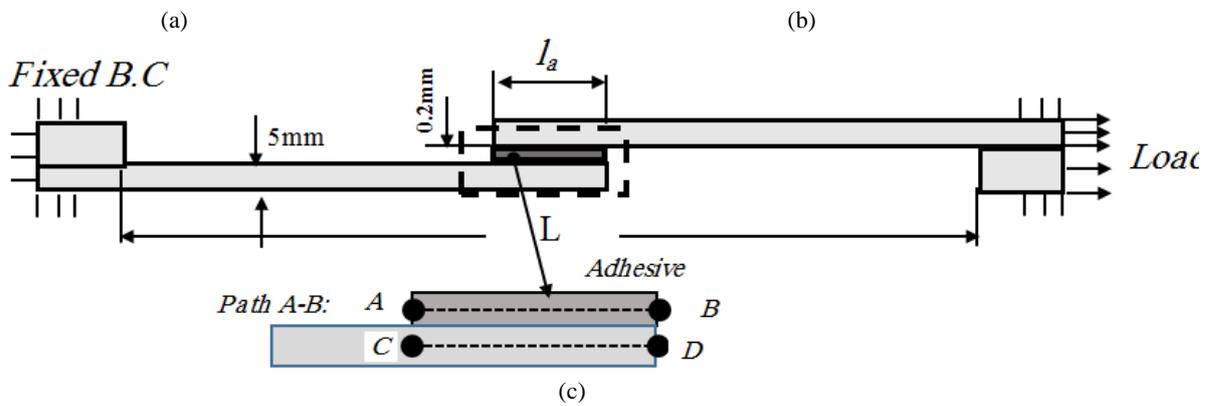

(c)

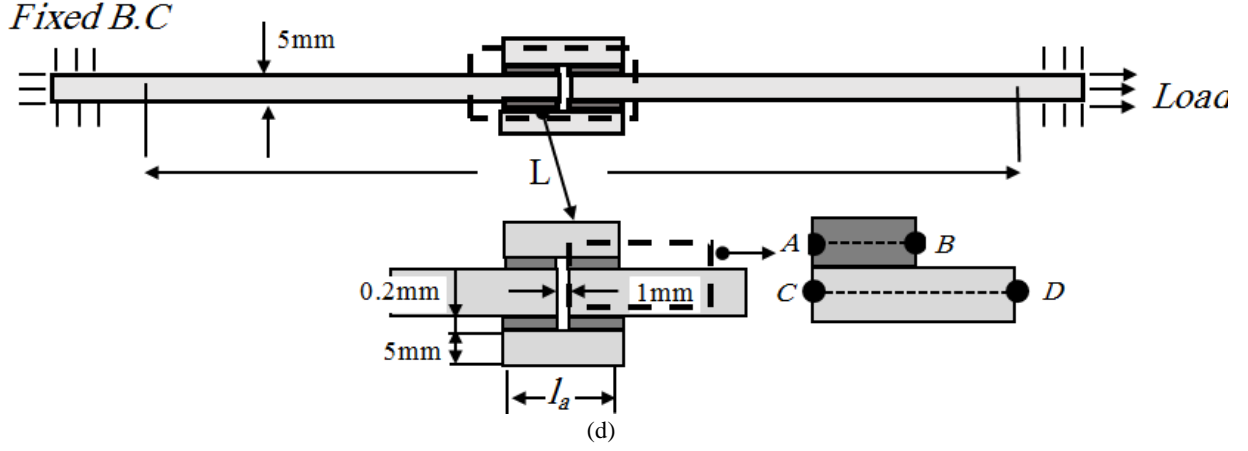

**Fig. 1.** Experimental equipment and sketches of the adhesive joint configurations: a) a SL joint under uniaxial tensile test, b) uniaxial creep test of a polyethylene bulk sample, c) SL joint, and d) DS joint.

## 3. Numerical model

In the following, we proceed to describe the constitutive models employed and the numerical framework that has been developed.

3.1 Constitutive models

The behaviour of the adherent and the adhesive is assumed to be elastic-plastic, in agreement with the tensile tests. J2 plasticity theory is used to model the adherent, as in [8]. The hardening behaviour follows the uniaxial stress-strain response, see Section 4. On the other hand, the adhesive is pressure-sensitive and consequently the Drucker–Prager model is used to constitutively characterise its behaviour [17]. Specifically, we adopt the general exponent form by which the yield function $F$ is written in the meridional plane ($p - q$ plane) as:

$$F = aq^b - p - p_t = 0 \qquad (1)$$

Where $p_t$ is the hardening parameter that represents the hydrostatic tension strength of the material, $p$ is the hydrostatic stress, $q$ is the effective stress, and $a$ and $b$ are material parameters. For a given material yield stress in tension $\sigma_Y$ and a hydrostatic-stress-sensitivity parameter $\lambda$, the values of $a$ and $p_t$ can be given as:

$$a = \frac{1}{3\sigma_Y(\lambda - 1)} \qquad (2)$$

$$p_t = a\lambda\sigma_Y^2 \qquad (3)$$

Plastic flow is defined by the parameter $\psi$ and is obtained from the following expression:

$$\tan\psi = \frac{3(1 - 2\nu^p)}{2(1 + \nu^p)} \qquad (4)$$

where $\nu^p$ is the plastic component of Poisson's ratio [17].

On the other hand, capturing the time-dependent behaviour of the adhesive joints requires modelling the creep behaviour of the adhesive, Araldite 2011, and the adherent, polyethylene. The experiments are conducted at a load level well below yielding, and accordingly the evolution of the creep strain as a function of time, $\varepsilon_c(t)$, is defined by subtracting the elastic strains $\varepsilon_{elastic}$ to the total strains $\varepsilon_{total}$:

$$\varepsilon_c(t) = \varepsilon_{total}(t) - \varepsilon_{elastic} \qquad (5)$$

The non-linear creep behaviour of polymers can be appropriately captured using a sufficient number of elastic and damping elements. Here, a model is used that combines spring and damper elements to derive the compliance equation of the material, as sketched in Fig. 2. This model is a combination of Zener and Maxwell models [18] and can capture both first and secondary creep stages. However, attention is here limited to the response of adhesive joints in the primary creep regime.

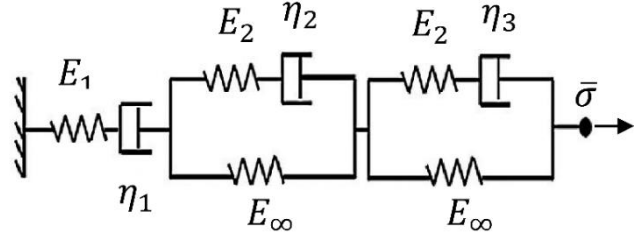

**Fig. 2.** Schematic representation of the combination of spring and damper elements used in the rheological model assumed, which is composed of one Maxwell and two Zener models.

The model compliance can be obtained by making use of Laplace's transform. For a given average stress $\bar{\sigma}$, the evolution in time of the total strain is a function of several material constitutive parameters, as:

$$\varepsilon_{total}(t) = \bar{\sigma}\left[\frac{1}{E_1} + \frac{t}{\eta_1} + \frac{2}{E_\infty} - \frac{E_0 - E_\infty}{E_0}e^{-\frac{t}{\theta_1}} - \frac{E_0 - E_\infty}{E_0}e^{-\frac{t}{\theta_2}}\right] \quad (6)$$

The material parameters ($E_\infty, E_1, E_2, \eta_1, \eta_2$ and $\eta_3$) can be obtained through nonlinear regression of the experimental results of creep tests. On the other hand, the parameters $E_0, \theta_1$ and $\theta_2$ are defined as:

$$E_0 = E_\infty + E_2; \quad \theta_1 = \frac{\eta_2 E_0}{[E_\infty(E_0 - E_\infty)]}; \quad \theta_2 = \frac{\eta_3 E_0}{[E_\infty(E_0 - E_\infty)]} \quad (7)$$

Equation (6) can be reformulated [19] such that the rheological model reads:

$$\varepsilon_{total}(t) = \varepsilon_e + \left[\alpha\hat{t} - \beta\left(e^{-a_5\hat{t}} + e^{a_5\hat{t}}\right)\right] = \varepsilon_e + \alpha\hat{t} + \beta\sinh(a_6\hat{t}) \quad (8)$$

Where $\varepsilon_e$ denotes the elastic strain and α and β are respectively referred to as the first and second-order functions of stress. The parameters α, β and $\hat{t}$ can be defined as follows:

$$\alpha = a_1 + a_2\bar{\sigma}; \quad \beta = a_3 + a_4\bar{\sigma} + a_5\bar{\sigma}^2; \quad \hat{t} = \log_{10}(t) \quad (9)$$

On the other hand, the parameters $a_i$ are material constants that can be calculated from the experimental data by means of nonlinear regression techniques. In the present work, the mathematical software MATLAB is used to exact these coefficients – see Section 4.

3.2 Numerical implementation

The commercial finite element package ABAQUS is used to reproduce the experimental campaign described in Section 2. Two-dimensional, plane stress models are developed to reproduce the SL and DS adhesive joint configurations shown in Figs. 1c and 1d. For the meshing, we use eight-node quadratic elements with reduced integration, CPS8R in ABAQUS notation. After a mesh sensitivity analysis, it is found that numerical convergence is achieved when using about 16000 and 2400 elements in, respectively, the adherent and the adhesive parts. A representative detail of the mesh is shown in Fig. 3 below. The boundary conditions employed mimic the experiments, as depicted in Figs. 1c and 1d. The same models are used for both the tensile and the creep tests. The constitutive model for creep behaviour described in Section 3.1 is implemented by means of a user material (UMAT) subroutine in ABAQUS.

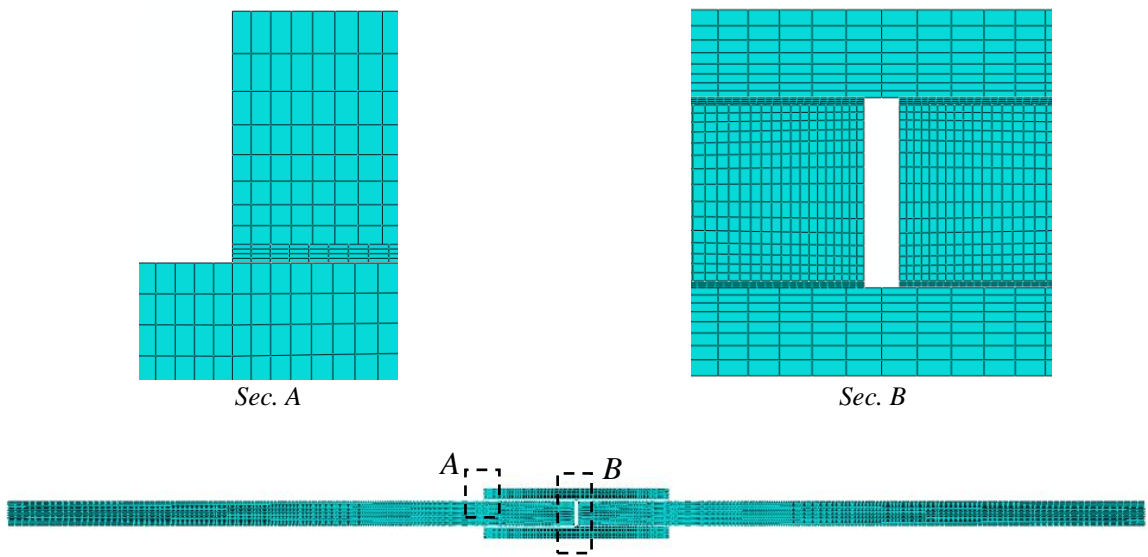

Fig. 3. General and detailed representation of the finite element mesh employed.

# 4. Results and discussion

In this section, we describe the experimental and numerical results obtained, as well as their implications on adhesive joint behaviour and design. Tensile testing on the adherent and adhesive materials is described first, Section 4.1, followed by their creep responses, Section 4.2. In Section 4.3, we address the tensile response of the adhesive joints predicted both experimentally and numerically. Then, in Section 4.4, we examine the creep responses of SL and DL adhesive joints and use the finite element model to gain further insight.

## 4.1 Tensile tests of the adherent and adhesive materials

Uniaxial tensile tests are conducted to characterise the mechanical response of the adherent, polyethylene. The mechanical properties of the polyethylene obtained from experiments are listed in Table 1, along with the mechanical properties of the adhesive, provided by the manufacturer.

**Table 1**: Mechanical properties of polyethylene and the epoxy-based adhesive Araldite 2011.

|  | Araldite 2011, [20] | Polyethylene |
|---|---|---|
| Young's modulus, E (MPa) | 1802 ± 20.1 | 1154 ± 15.4 |
| Poisson's ratio, $\upsilon$ | 0.29 ± 0.04 | 0.3 ± 0.2 |
| Tensile yield strength, $\sigma_y$ (MPa) | 18 ± 0.6 | 14.02 ± 0.3 |
| Tensile failure strength, $\sigma_f$ (MPa) | 26.36 ± 0.48 | 20.01 ± 0.25 |

A representative stress-strain curve for polyethylene is shown in Fig. 4. This stress-strain hardening behaviour is provided as input to the finite element model. Recall that J2 plasticity theory is used for polyethylene while the exponential form of Drucker-Prager is employed for the adhesive. The Drucker-Prager parameters used for Araldite 2011 are $a = 0.092$, $b = 2$ and dilatation angle $\psi = 13°$, as reported elsewhere [3,21].

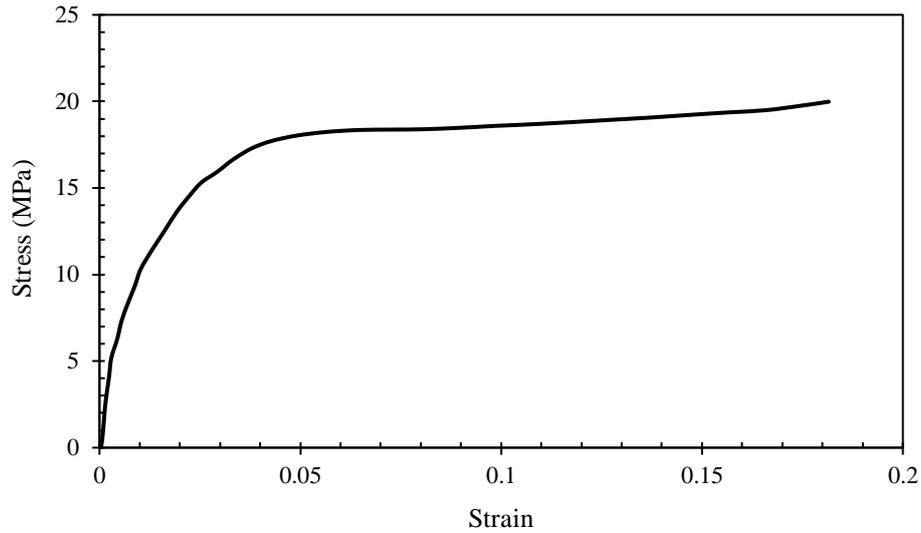

**Fig. 4.** The uniaxial stress-strain curve obtained from testing polyethylene.

4.2 Creep tests of the adherent and adhesive materials

We proceed to characterise the creep behaviour of the adhesive material, Araldite 2011, and the adherent material, polyethylene. The experiments are conducted under constant load and uniaxial tension conditions. Results obtained for the adhesive material are reported in Fig. 5, in terms of creep strain versus time (in hours). Four selected values of the remote stress are considered: 15.3 MPa, 13.5 MPa, 11.7 MPa and 9.9 MPa. In agreement with expectations, the creep strain increases with the applied stress.

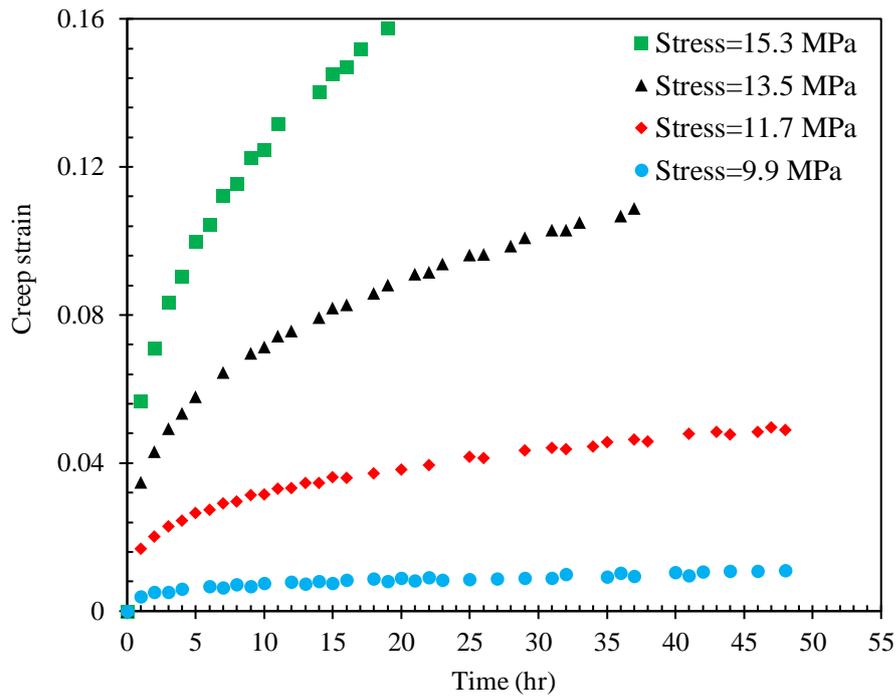

**Fig. 5.** Uniaxial creep test results for the adhesive material, Araldite 2011.

The results obtained for the adherent material, polyethylene, are shown in Fig. 6. Four remote load levels are considered, as characterised by remote stresses of 12 MPa, 10.5 MPa, 9 MPa and 7.5 MPa. As for the adhesive, the creep strain naturally increases with the applied stress.

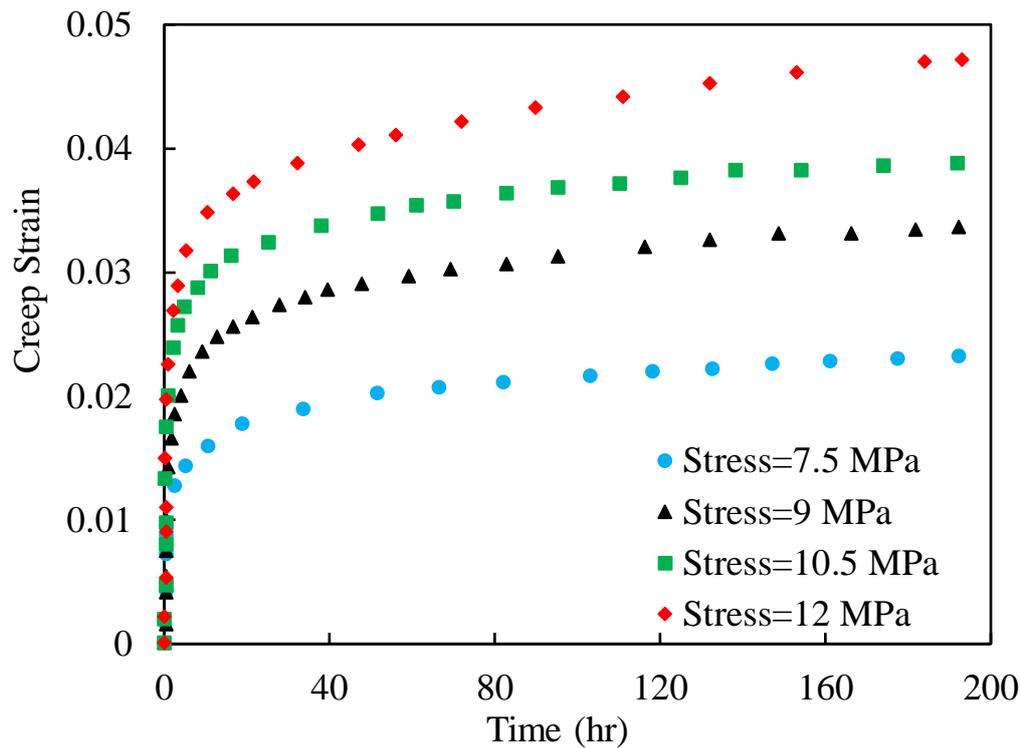

**Fig. 6.** Uniaxial creep test results for the adherent material, polyethylene.

The experimental data shown in Figs. 5 and 6 are used to calibrate the creep constitutive model outlined in Section 3.1. The software MATLAB is used to obtain the corresponding coefficients using nonlinear regression. For both the adhesive and adherent materials the coefficient of determination $R^2$ is very close to 1. The parameters obtained are listed in Table 2. The data was then used to model the creep behaviour of the adhesive joint systems.

Table 2: Coefficients of creep constitutive model for the adhesive and the adherent materials.

|  | $a_1$ | $a_2$ | $a_3$ | $a_4$ | $a_5$ | $a_6$ | $R^2$ |
|---|---|---|---|---|---|---|---|
| Adhesive | 1.59E-02 | -1.41E-03 | 5.50E-03 | -2.28E-03 | 2.28E-04 | 0.92 | 0.991 |
| Polyethylene | -2.06E-02 | 2.99E-04 | 9.10E-02 | 1.41E-02 | -5.02E-04 | 1.22E-01 | 0.989 |

4.3 Modelling and testing of the tensile behaviour of adhesive joints

Once the individual constitutive behaviour of the adherent and the adhesive have been characterised, we proceed to model and test the behaviour of the SL and DS joints. The load versus displacement curves measured from three sets of experiments are shown in Figs. 7a and 7b for, respectively, SL and DS joints. The case of an overlap length $l_a$=19 mm is considered. The three experiments conducted for each joint configuration show a good degree of reproducibility. In addition, the numerical prediction is also shown. The finite element results exhibit good agreement with the experiments in both the linear and non-linear regime, and for the two joint configurations considered. The agreement attained validates the constitutive modelling framework adopted.

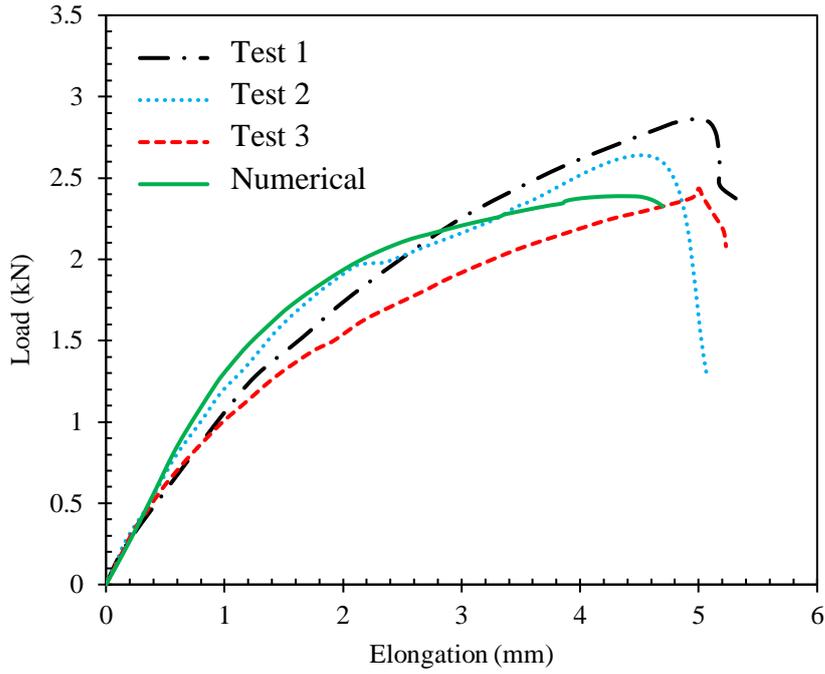

(a)

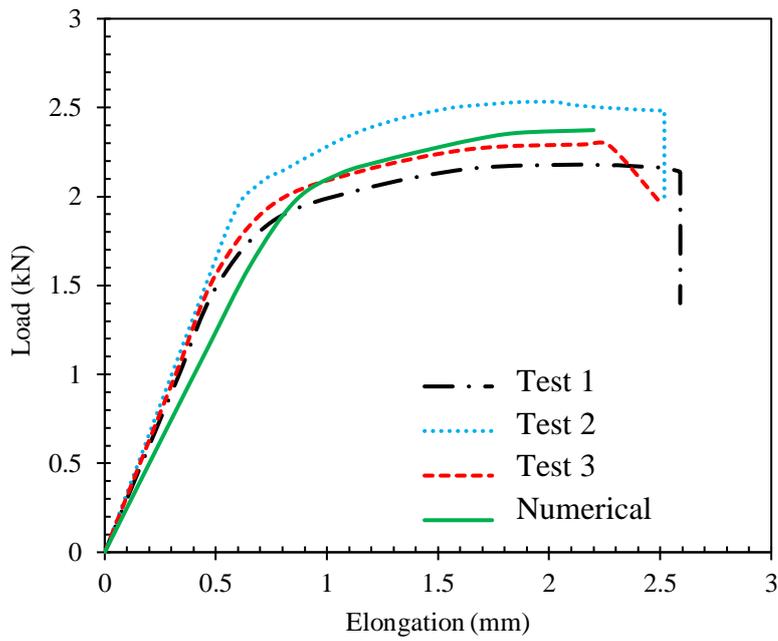

(b)

**Fig. 7.** Load versus displacement curves obtained experimentally (3 tests) and numerically for adhesive joints with overlap length of $l_a$=19 mm and subjected to uniaxial tension; a) SL adhesive joints, and b) DS adhesive joints.

Uniaxial tests were also conducted on SL and DS joints with overlap lengths of $l_a$=29 and 39 mm. The results obtained, in terms of the failure load, are shown in Fig. 8. We find that the failure loads are sensitive to the overlap length, with larger overlap lengths leading to higher failure loads. Also, SL joints appear to show a better performance, failing at higher load levels, relative to DS joints. The reason could be the higher number of stress concentration points in the DS geometry, relative to the SL joints.

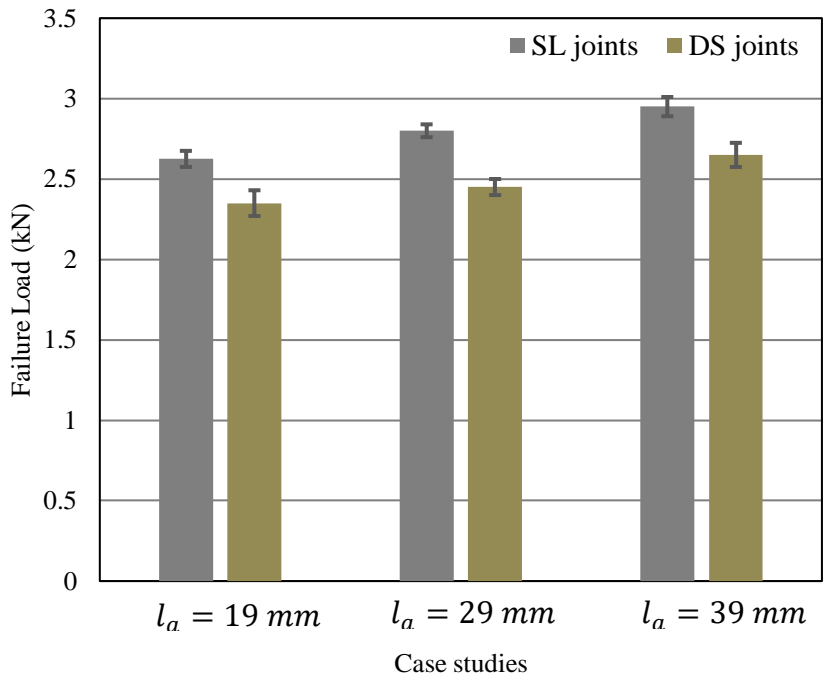

**Fig. 8.** Failure loads reported for SL and DS adhesive joint configurations and different overlap lengths $l_a$.

The validated model is employed to assist in the design of the creep experiments. The goal is to ensure that the applied load is sufficiently low such that no yielding occurs in the adhesive and, to a certain extent, in the adherent. First, the von Mises equivalent stress is plotted along the centre line of the adhesive, path A-B in Figs. 1c and 1d. The results are shown in Fig. 9 for a remote load of 2 kN and both SL and DS joints with different overlap lengths. The effective von Mises stress is shown versus the distance along the centre line, which is normalised by the overlap length.

It is shown that a load of 2 kN is the highest that can be considered without triggering yielding close to the edges of the adhesive. Recall, Table 1, that the adhesive yield strength equals 18 MPa, such that a further increase in the remote yielding will lead to effective von Mises stresses larger than this value at the edges. On the other hand, the adherent has a yield strength of 14 MPa. Similar trends are observed for the stress distribution in both SL and DS joints and, in all cases, smaller overlap lengths lead to higher stress values.

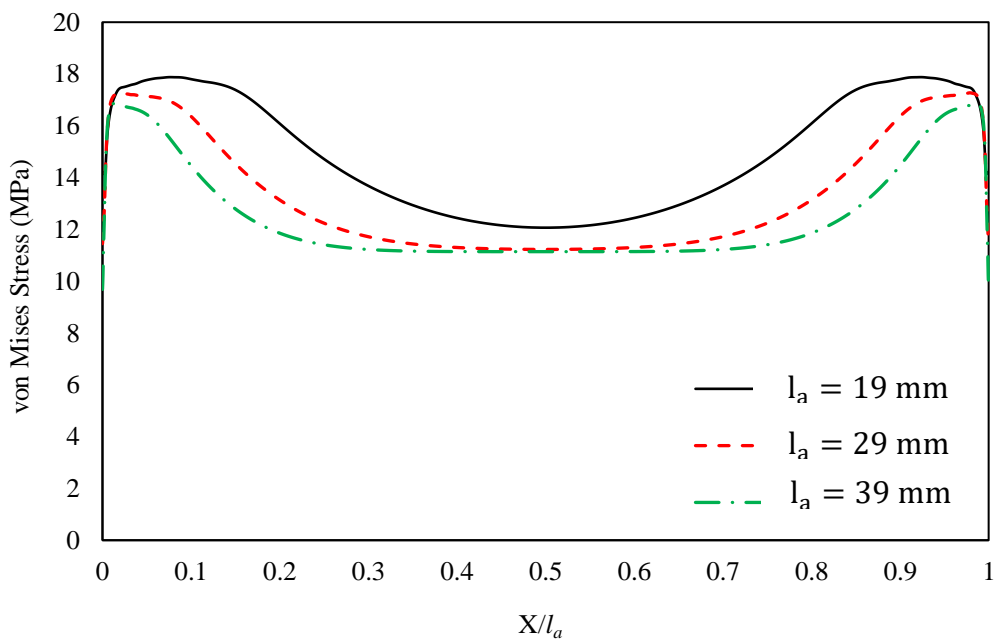

(a)

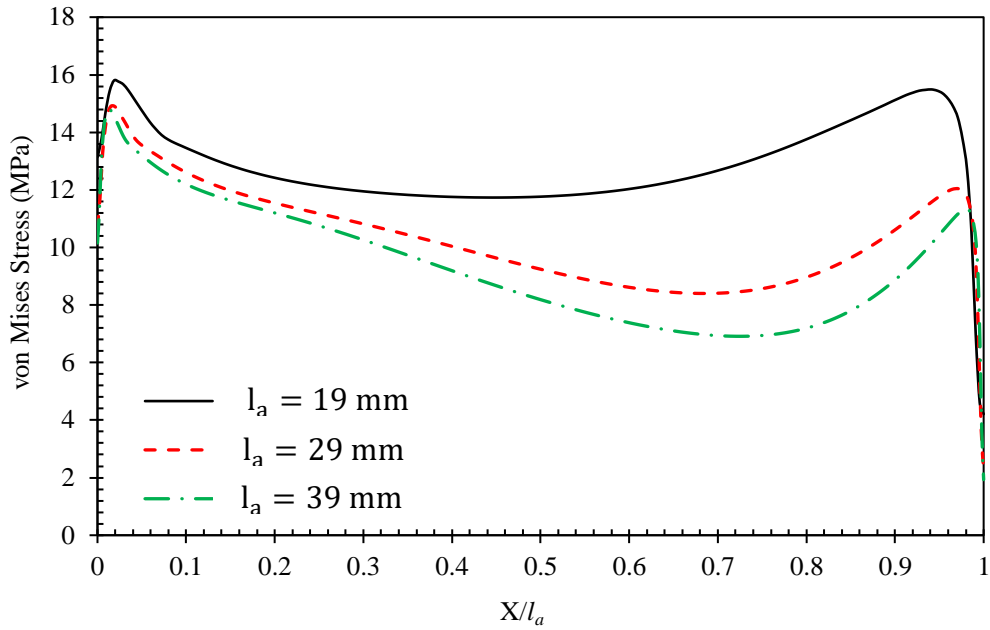

(b)

**Fig. 9**. Distribution of the equivalent von Mises stress along the centre line of the adhesive; (a) SL joint and (b) DS joint. The $x$ axis shows the distance along AB path normalised by the overlap length $l_a$ (see Fig. 1).

Secondly, the equivalent von Mises stress is also computed in the adherent material. The contours of equivalent von Mises stress are plotted in Fig. 10 for both SL and DS joint configurations; the units are MPa. The case of overlap length $l_a$=19 mm and remote load equal to 2 kN is chosen as representative. While the equivalent von Mises stress exceeds the yield stress in some small regions, it remains below overall and particularly underneath the overlap region. The yielding areas can undoubtedly influence the outcome of the experiment *via* adherent distortion.

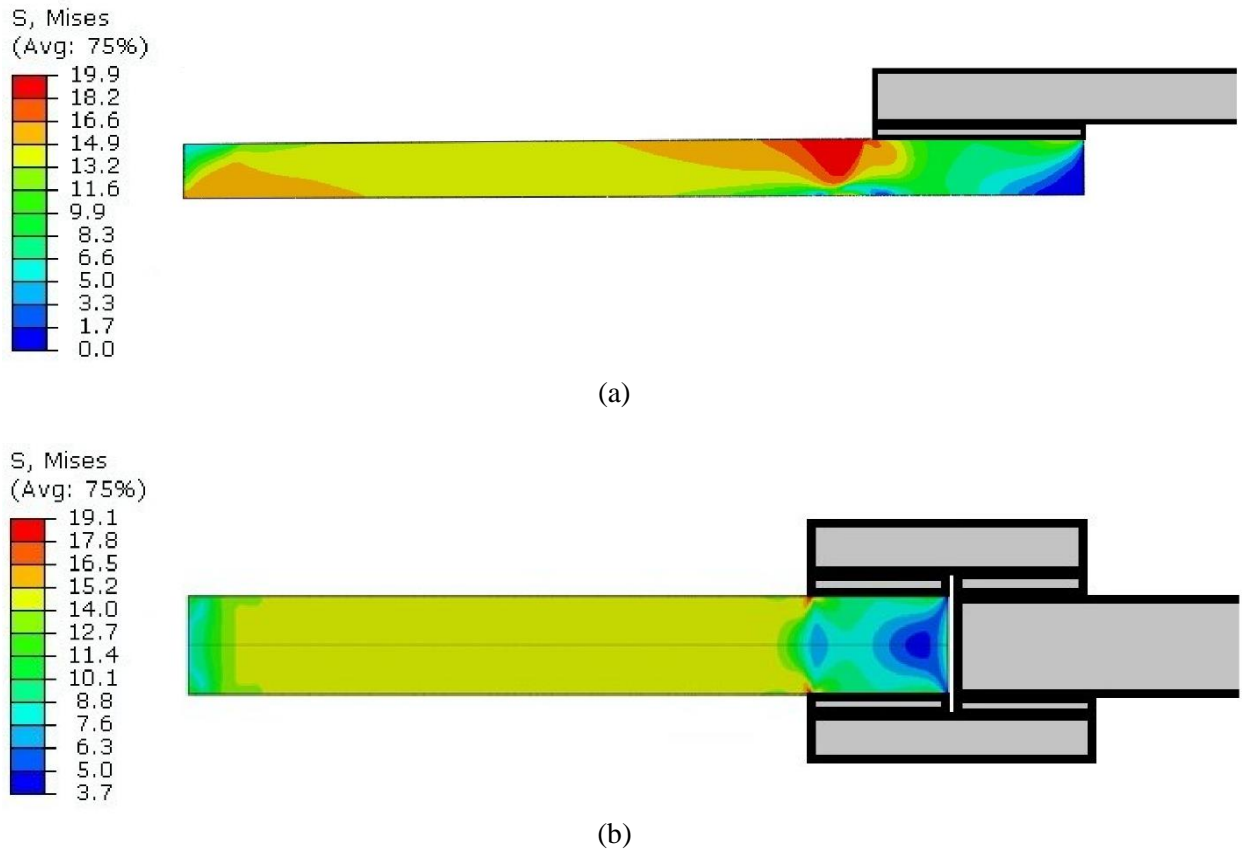

**Fig. 10.** Contours of von Mises stress (MPa) in the adherend material; (a) SL and (b) DS joint configurations.

4.4 Modelling and testing of the creep behaviour of adhesive joints

Uniaxial creep tests are conducted on the adhesive joints following the results obtained in Section 4.3. Two adhesive joint configurations are considered, SL and DS and, for each configuration, we vary the overlap length $l_a$, as for the uniaxial monotonic tests. Six adhesive joint configurations are considered, and each of them is subjected to two remote loads: 1.6 and 2 kN; a total of 12 case studies (see Fig. 11). The magnitude of the remote load is chosen based on the numerical analysis, aiming to minimise yielding in the adhesive and adherent materials. The experiments are conducted for 83 hours to characterise the primary creep regime (steady creep is observed after this time). None of the samples fails during this time. The results obtained for each case study are shown in Fig. 11 in terms of the elongation versus the testing time. In addition, the numerical predictions obtained with the model presented in Section 4, and validated as described above, are also shown. Symbols denote experimental results while solid lines are used to describe the finite element predictions. The numerical results slightly underpredict the experimental elongation-time responses but the agreement is satisfactory.

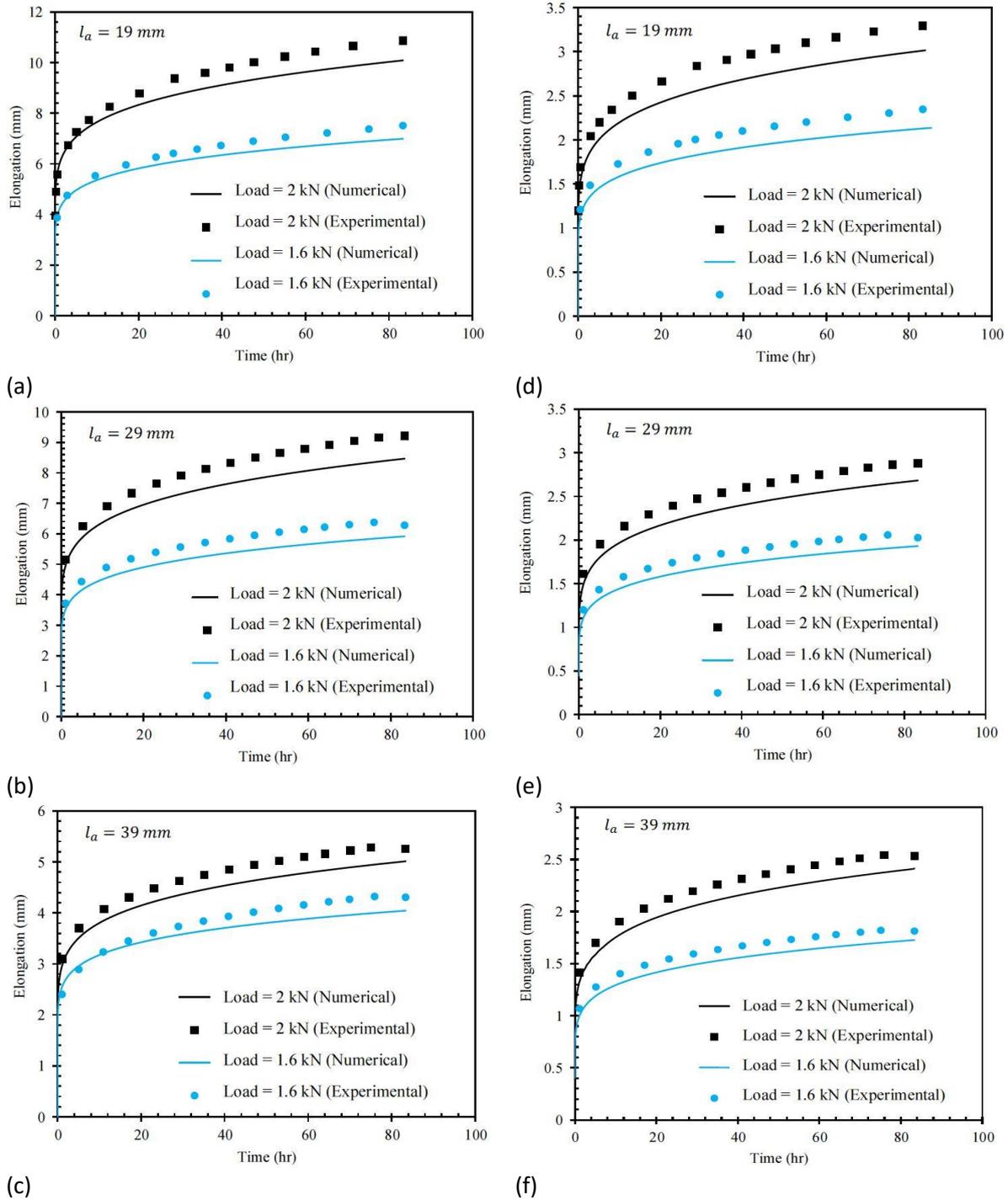

**Fig. 11.** Experimental and numerical time-elongation responses under creep loading of SL joints (a): $l_a$=19 mm (b): $l_a$=29 mm (c): $l_a$=39 mm and DS joints (d): $l_a$=19 mm (e): $l_a$=29 mm (f): $l_a$=39 mm.

In both the experimental and numerical predictions, the following trends can be observed. First, in both SL and DS joints, the elongation increases with decreasing the overlap length. Moreover, in agreement with expectations, the elongation increases with the applied load. Thirdly, significantly higher elongations are observed in the SL joints, relative to the DS configuration. For example, for $l_a$=19 mm and a load of 2 kN, the maximum elongation recorded for the SL joints is more than three times the DS measurement; 10 mm and 3 mm, respectively. These differences diminish with increasing overlap length but remain substantial in all cases; for an overlap length of $l_a$=39 mm the elongation of SL joints is approximately twice of that predicted by DS joints for the same remote load. This trend can be explained by comparing the geometry of the joints. Although the overlap length is kept equal, the bonding length in DS joints is approximately two times the bonding length of the SL joints. This difference raises the DS joints' stiffness, resulting in lower deformations. Also, note the higher degree of yielding predicted in the adherent for the SL case (Fig. 10).

We proceed to gain further insight into the creep response by making use of the numerical model. Specifically, we aim at quantifying the stress redistribution that occurs during the testing. First, we compute the peel and shear stresses along the centre line of the adhesive, referred to as path A-B in Figs. 1c and 1d. The stress distributions are calculated at the end of the loading step ($t \approx 0$) and the end of the creep test ($t = 83$ h). The results are shown in Fig. 12 for the representative case of an overlap length $l_a = 19\ mm$, and for both SL and DS adhesive joint configurations.

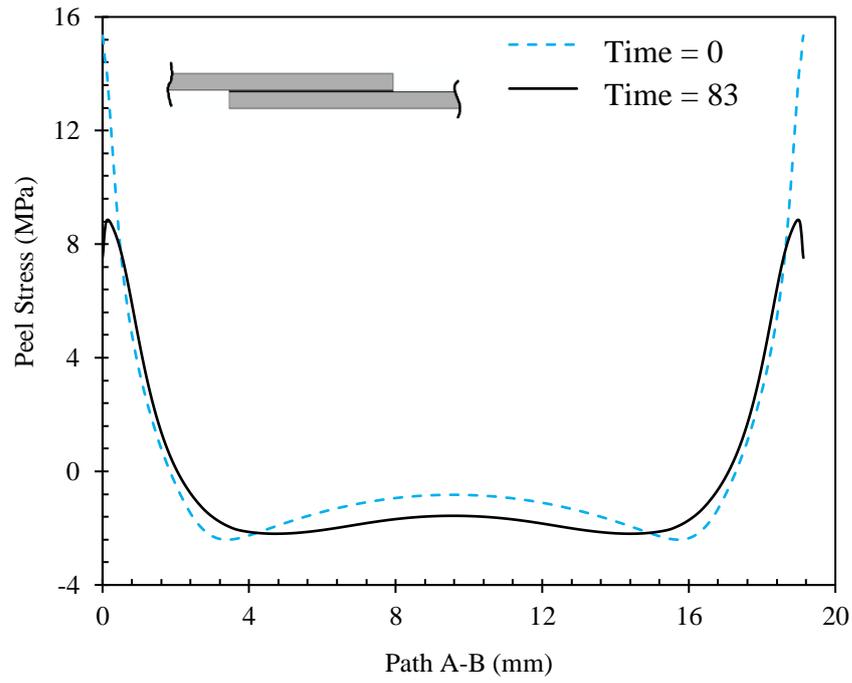

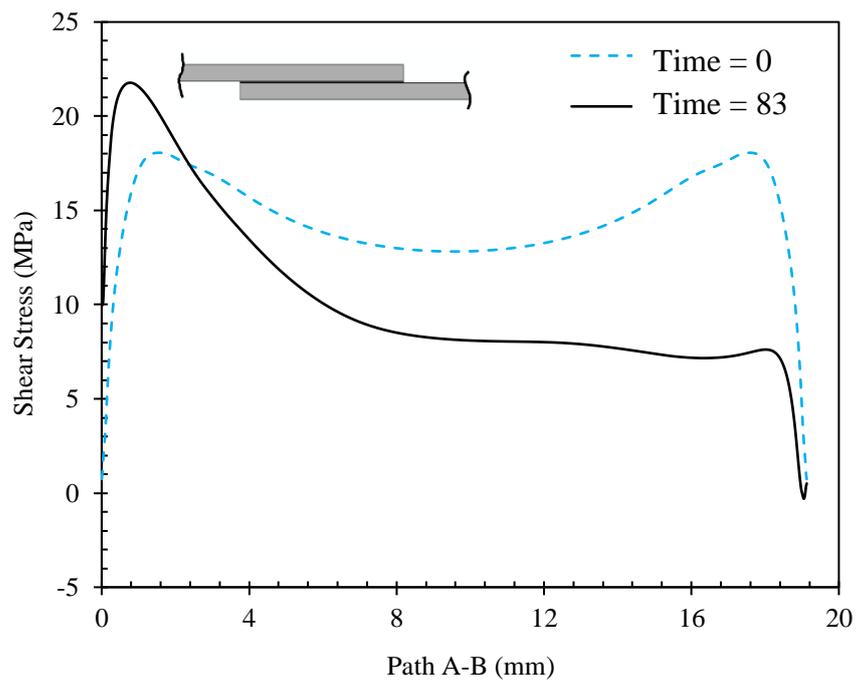

(a)

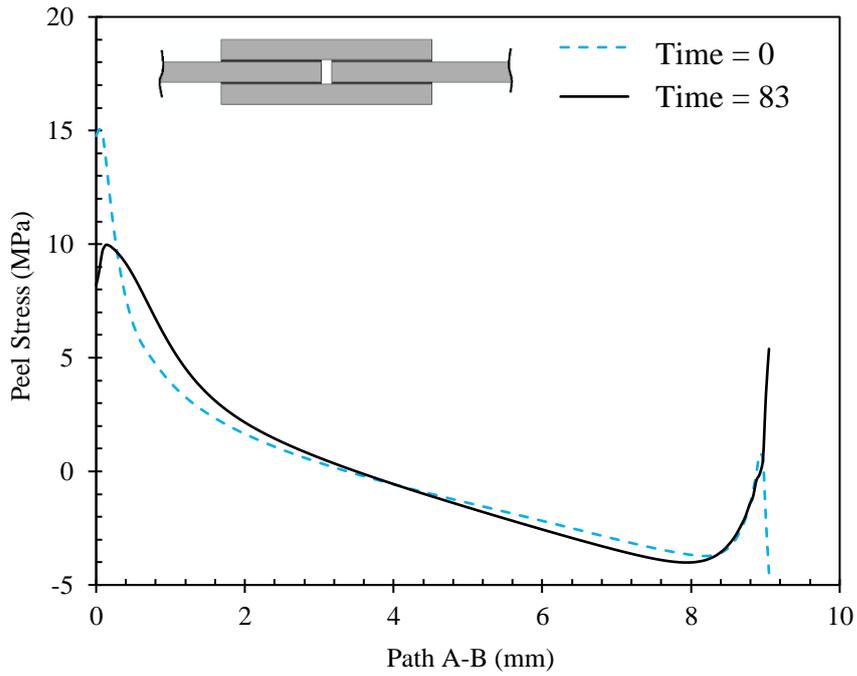

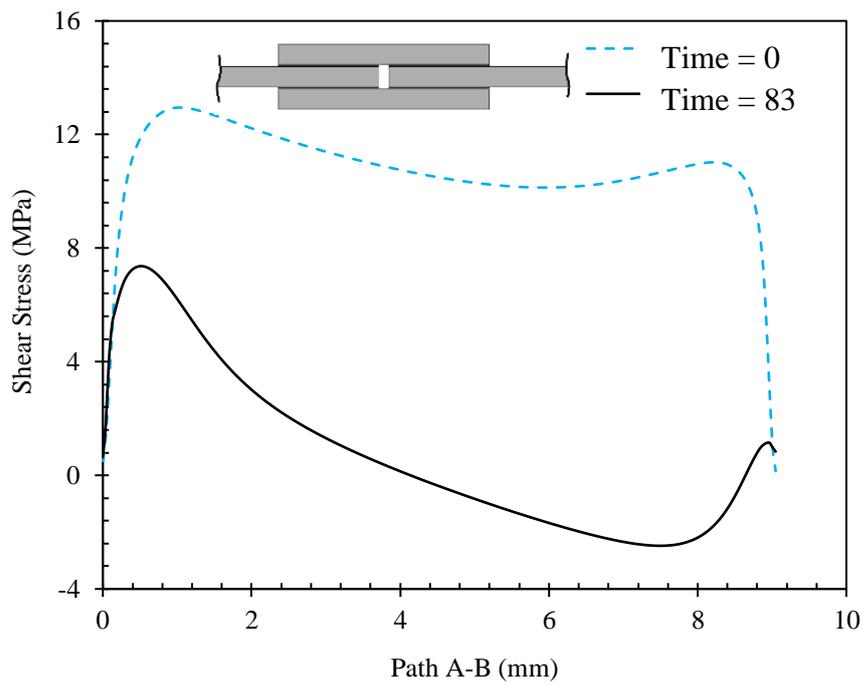

**(b)**

**Fig. 12** Stress redistribution along the centre-line of the adhesive, path A-B. The result at the end of the loading step ($t \approx 0$) is given by a dashed line, while the result at the end of the creep test ($t \approx 83$ h) is given by a solid line. (a) SL, and (b) DS joints.

A significant peel stress reduction is observed at the overlap end regions, however farther away the stress sensitivity to creep is diminished. Changes are more substantial in the shear case. In both SL and DS joints the shear stress level decreases with the testing time at almost every point of the adhesive layer. The drop is particularly significant in the case of the DS joint configuration, where negative shear stresses are attained in a region of the path.

We also explore the stress state in the adherent, see Fig. 13. The effective von Mises stress is plotted along the centre line of the adherent (path C-D in Fig. 1), polyethylene, at both the end of the loading step ($t \approx 0$) and the end of the creep test ($t = 83$ h). Again, both SL, Fig. 13a, and DS, Fig. 13b configurations are considered. Overall, a small stress redistribution is observed. In this case, differences with the initial state are more noticeable for the SL joint configuration, with relevant stress changes being observed at the edge of the path.

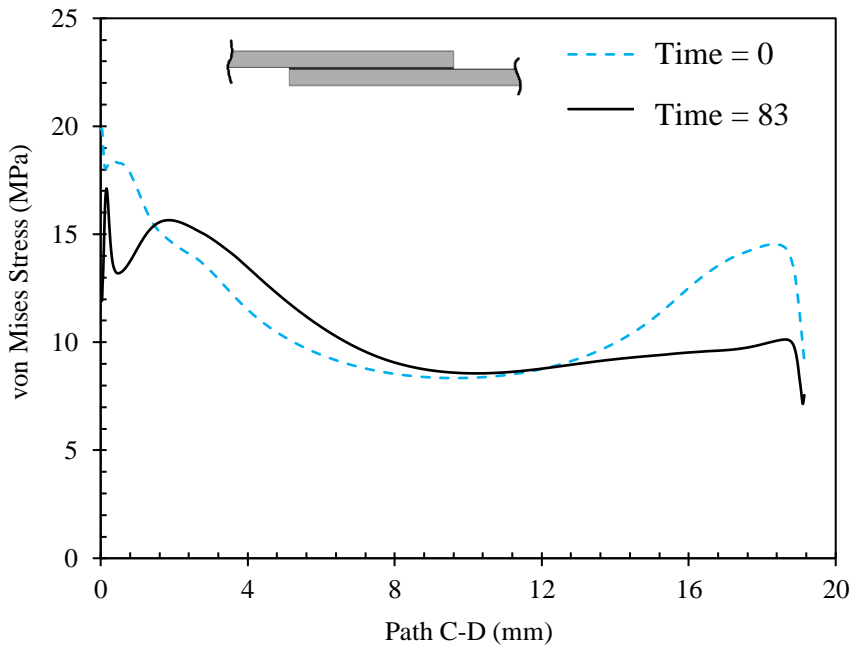

(a)

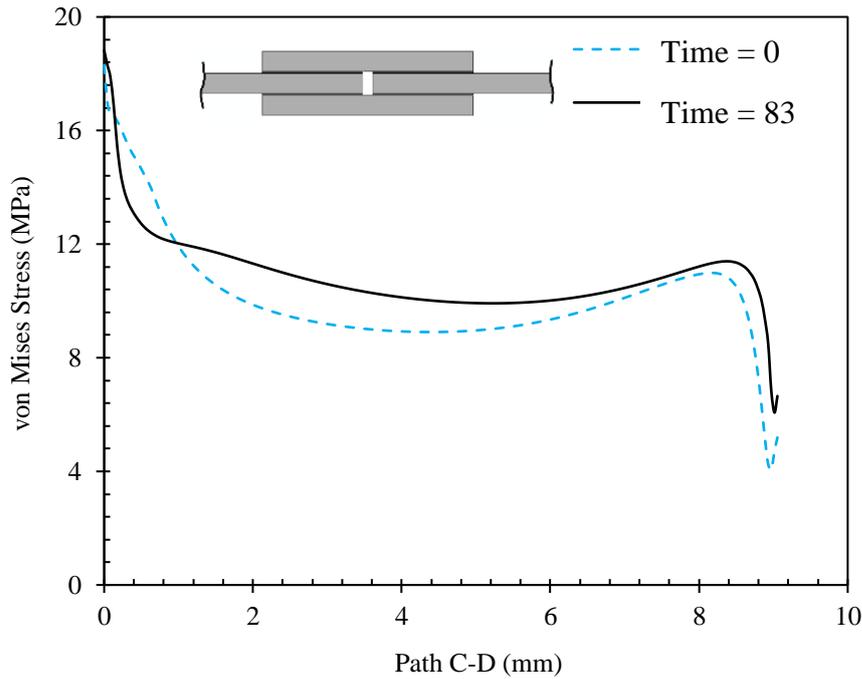

(b)

**Fig.13** Stress redistribution along the centre-line of the adherent, path C-D. The result at the end of the loading step ($t \approx 0$) is given by a dashed line, while the result at the end of the creep test ($t \approx 83$ h) is given by a solid line. (a) SL, and (b) DS joints. The case of overlap length $l_a = 19\ mm$ is taken as reference.

## 5. Conclusion

We investigated, numerically and experimentally, the static and creep behaviour of polyethylene-based Single-Lap (SL) and Double-Strap (DS) adhesive joints. First, insight is gained into the behaviour of the adherent and adhesive bulk materials, and a suitable rheological model is developed to capture their time-dependent response. The model is implemented in the finite element package ABAQUS by programming a user material subroutine. A large experimental campaign is conducted to evaluate the behaviour of SL and DS joints under static and time-dependent conditions for different load levels and overlap lengths. By combining experiments and numerical simulations we reveal the following findings:

- SL joints slightly outperform DS joints under uniaxial tension. In both systems, the failure load increases with increasing overlap length.

- Under creep conditions and for a given remote load, SL joints reveal much larger elongations than DS joints. Differences decrease with increasing overlap length but remain substantial in all cases.

- Little redistribution of the peel stress is observed in the adhesive. However, the shear stress shows a notable sensitivity to the testing time, and this effect is more pronounced for the DS joint configuration.

The numerical model shows a very good agreement with the experiments, strengthening the constitutive choices and enabling the assessment of multiple configurations for optimising adhesive joint design. The qualitative and quantitative insight gained into the mechanical and creep behaviours of polyethylene-based joints should facilitate design and industrial uptake.